\documentclass[11pt]{article}\textwidth 6.5in\textheight 9in
\usepackage{amssymb}\usepackage[colorlinks]{hyperref}\usepackage{color}
	\usepackage{stmaryrd}\usepackage{mathrsfs}
		\usepackage{graphicx}
	\usepackage{amsmath}
\usepackage{caption}
    \usepackage{subcaption}
    \usepackage{listings}
 
\begin{document}
\setlength{\captionmargin}{27pt}
\newcommand\hreff[1]{\href {http://#1} {\small http://#1}}
\newcommand\trm[1]{{\bf\em #1}} \newcommand\emm[1]{{\ensuremath{#1}}}
\newcommand\prf{\paragraph{Proof.}}\newcommand\qed{\hfill\emm\blacksquare}

\newtheorem{thr}{Theorem} 
\newtheorem{lmm}{Lemma}
\newtheorem{cor}{Corollary}
\newtheorem{con}{Conjecture} 
\newtheorem{prp}{Proposition}

\newtheorem{blk}{Block}
\newtheorem{dff}{Definition}
\newtheorem{asm}{Assumption}
\newtheorem{rmk}{Remark}
\newtheorem{clm}{Claim}
\newtheorem{exm}{Example}

\newcommand\Ks{\mathbf{Ks}} 
\newcommand{\ab}{a\!b}
\newcommand{\yx}{y\!x}
\newcommand{\yux}{y\!\underline{x}}

\newcommand\floor[1]{{\lfloor#1\rfloor}}\newcommand\ceil[1]{{\lceil#1\rceil}}

\newcommand{\lea}{<^+}
\newcommand{\gea}{>^+}
\newcommand{\eqa}{=^+}

\newcommand{\lel}{<^{\log}}
\newcommand{\gel}{>^{\log}}
\newcommand{\eql}{=^{\log}}

\newcommand{\lem}{\stackrel{\ast}{<}}
\newcommand{\gem}{\stackrel{\ast}{>}}
\newcommand{\eqm}{\stackrel{\ast}{=}}

\newcommand\edf{{\,\stackrel{\mbox{\tiny def}}=\,}}
\newcommand\edl{{\,\stackrel{\mbox{\tiny def}}\leq\,}}
\newcommand\then{\Rightarrow}

\newcommand\C{\mathbf{C}} 

\renewcommand\chi{\mathcal{H}}
\newcommand\km{{\mathbf {km}}}\renewcommand\t{{\mathbf {t}}}
\newcommand\KM{{\mathbf {KM}}}\newcommand\m{{\mathbf {m}}}
\newcommand\md{{\mathbf {m}_{\mathbf{d}}}}\newcommand\mT{{\mathbf {m}_{\mathbf{T}}}}
\newcommand\K{{\mathbf K}} \newcommand\I{{\mathbf I}}

\newcommand\II{\hat{\mathbf I}}
\newcommand\Kd{{\mathbf{Kd}}} \newcommand\KT{{\mathbf{KT}}} 
\renewcommand\d{{\mathbf d}} 
\newcommand\D{{\mathbf D}}

\newcommand\w{{\mathbf w}}
\newcommand\Cs{\mathbf{Cs}} \newcommand\q{{\mathbf q}}
\newcommand\E{{\mathbf E}} \newcommand\St{{\mathbf S}}
\newcommand\M{{\mathbf M}}\newcommand\Q{{\mathbf Q}}
\newcommand\ch{{\mathcal H}} \renewcommand\l{\tau}
\newcommand\tb{{\mathbf t}} \renewcommand\L{{\mathbf L}}
\newcommand\bb{{\mathbf {bb}}}\newcommand\Km{{\mathbf {Km}}}
\renewcommand\q{{\mathbf q}}\newcommand\J{{\mathbf J}}
\newcommand\z{\mathbf{z}}

\newcommand\B{\mathbf{bb}}\newcommand\f{\mathbf{f}}
\newcommand\hd{\mathbf{0'}} \newcommand\T{{\mathbf T}}
\newcommand\R{\mathbb{R}}\renewcommand\Q{\mathbb{Q}}
\newcommand\N{\mathbb{N}}\newcommand\BT{\Sigma}
\newcommand\FS{\BT^*}\newcommand\IS{\BT^\infty}
\newcommand\FIS{\BT^{*\infty}}
\renewcommand\S{\mathcal{C}}\newcommand\ST{\mathcal{S}}
\newcommand\UM{\nu_0}\newcommand\EN{\mathcal{W}}

\newcommand{\supp}{\mathrm{Supp}}

\newcommand\lenum{\lbrack\!\lbrack}
\newcommand\renum{\rbrack\!\rbrack}

\newcommand\h{\mathbf{h}}
\renewcommand\qed{\hfill\emm\square}
\renewcommand\i{\mathbf{i}}
\newcommand\p{\mathbf{p}}
\renewcommand\q{\mathbf{q}}
\title{ Derandomization under Different Resource Constraints}

\author {Samuel Epstein\footnote{JP Theory Group. samepst@jptheorygroup.org}}

\maketitle
\begin{abstract}
We provide another proof to the EL Theorem. We show the tradeoff between compressibility of codebooks and their communication capacity. A resource bounded version of the EL Theorem is proven. This is used to prove three instances of resource bounded derandomization. This paper is in support of the general claim that if the existence of an object can be proven with the probabilistic method, then bounds on its Kolmogorov complexity can be proven as well.
\end{abstract}
\section{Introduction}

A problem is a collection of instances and the goal is to determine whether the instance satisfies a certain property, i.e. has a solution. This can be formulized by a computable function $V:\FS\times\FS\rightarrow\BT$, where instances $x\in\FS$ are the first argument and solutions $y\in\FS$ are the second argument, i.e $\{y:V(x,y)=1\}$. An example is {\sc Sat}, where instances are formulas and the solutions are assignment of the variables which satisfies it. Through the recently introduced method of \textit{derandomization}, \cite{EpsteinDerandom22,EpsteinExmDerand22}, this approach can be aligned with Algorithmic Information Theory in a new way: bounds on the Kolmogorov complexity of the simplest solutions can be proven.  The notion of a “solution” is fluid, for example in {\sc MaxSat}, in \cite{EpsteinExmDerand22}, a solution could entail any assignment that satisfies 6/7 the optimal number of possible satisfiable clauses. 

The procedure for derandomization is as follows. The first step is to prove solutions to  certain instances of problems have solutions that occur with probability at least $p$, with respect to a simple probability measure P over the solution candidate space. This is done by often employing the Lov\'{a}sz Local Lemma (Lemma \ref{lmm:lll}). Then, by applying conservation of information (Lemma \ref{lmm:consH}) and the EL Theorem (Corollary \ref{cor:el}), bounds on the Kolmogorov complexity of the simpliest solution can be proven. More specifically there exists some solution encoded into $x\in\FS$, with
$$\K(x) \lel \K(P) -\log p + \I(\langle\textrm{description of the instance}\rangle;\chi).$$ 
$\K$ is the prefix-free Kolmogorov complexity function. $\I(x;\ch)=\K(x)-\K(x|\ch)$ is the asymmetric mutual information term between $x\in\FS$ and the halting sequence $\chi\in\IS$. The instance itself can be incredibly complex (for example a formula with exponential number of clauses to variables), but for all non-exotic instances, the term $ \I(\langle\textrm{description of the instance}\rangle;\chi)$ will be negligible. The main step of derandomization is the application of EL Theorem, also known as the Sets Have Simple Members Theorem:\\

\noindent\textbf{Theorem. (EL)} \textit{For finite $D\subset\FS, \min_{x\in D}\K(x)\lel \m(D)+\I(D;\ch)$.}\\
$ $\newpage

The term $\m(D)$ is equal to $\sum_{x\in D}\m(x)$, where $\m$ is the algorithmic probability. There are several proofs in the literature for the EL Theorem \cite{Shen12,Epstein19,Levin16}. In Section \ref{sec:el} of this paper we provide a new proof, which follows analogously to the proof in \cite{Levin16}, except left-total machines are not used.

In this paper, we will be applying derandomization to classical channels. Using derandomization, we will show a tradeoff between the compression size of a codebook vs. the communication capacity that the codebook allows. If the codebook is allowed more bits to be compressed to, the more capacity the codebook has in communicating information.  Derandomization of codebooks is possible because they can be proven to exist in classical information theory by using the probabilistic method. This paper is in support of the following claim.
\begin{clm}
If the existence of an object can be proven with the probabilistic method, then bounds on its Kolmogorov complexity can be proven as well.
\end{clm}

\subsection{Resource Bounded Derandomization}

In this paper, we show a resource bounded version of derandomization. By assuming the verifier $V$, defined in the previous section, runs in polynomial time, derandomizations can be reformulated using time bounded Kolmogorov complexity. To accomplish this, we introduce a resource bounded EL theorem, Corollary \ref{cor:resel}. This theorem follows almost directly from Theorem 4.1 in \cite{AntunesFo09}. The theorem is as follows. Let $\mathbf{FP}'=\{f:f\in\mathbf{FP}\textrm{ and } \|x\|=\|f(x)\|\}$.\\

\noindent\textbf{Theorem.} \textbf{(Resource Bounded EL)}  \textit{Assume \textbf{Crypto}. Let $L\in\mathbf{P}$, $A\in\mathbf{FP}'$, and assume $\delta_n=|\BT^n\cap A^{-1}(L_n)|/2^n$. Then for some polynomial $p$, $\min_{x\in L_n}\mathbf{K}^p(x) < -\log\delta_n+ O(\log n).$}\\

The function $\K^t$ is the $t$-time bounded Kolmogorov complexity and its formal definition can be found in Section \ref{sec:conv}. \textbf{Crypto} is a cryptographic assumption which ensures the existence of a certain type of pseudorandom generator. It can be found in Assumption \ref{asm}. This theorem enables resource bounded derandomization, in which certain problems constructed in uniform polynomial time have simple solutions, with respect to $\K^t$. The following theorem is, to our knowledge, the first of its kind.\\

\noindent\textbf{Theorem. (Resource Bounded Derandomization)} \textit{Assume \textbf{Crypto}. then}
\begin{enumerate}
\item \textit{Let $\{G_n\}$ be a uniformly computable in polynomial time sequence of $k$-regular graphs, with $k\geq 5$. There is a polynomial $p$ where for each $G_n$, there is a partition $x$ of $\lfloor\frac{k}{3\ln k}\rfloor$ components each containing a cycle with}
$$
\mathbf{K}^p(x) < 2n/k^2+O(\log n).
$$
\item \textit{For vector $v$, $\|v\|_\infty =\max_i |v_i|$. A binary matrix $M$ has entries of 0s or 1s. Let $\{M_n\}$ be a uniformly polynomial time computable sequence of $n\times n$ binary matrices. There is a polynomial $p$ where for each $M_n$ there is a vector $b\in \{-1,1\}^n$ such that $\|M_nb\|_\infty\leq 4\sqrt{n\ln n}$ and}
$$
\K^p(b) = O(\log n).
$$
\item \textit{Let $\Phi_n$ be a $k(n)$-SAT formula, using $n$ variables, $m(n)$ clauses, uniformly polynomial time computable in $n$. Furthermore, each variable occurs in at most $2^{k(n)}/k(n)e-1$ clauses. There is a polynomial $p$ and a satisfying assignment $x$ of $\Phi_n$ where} $$\mathbf{K}^p(x) < 2m(n)e2^{-k(n)} + O(\log n).$$
\end{enumerate}
\subsection{Future Work}
Future work entails exploring EL Theorems and Derandomizations under different resource constaints and access to random bits. For example if the universal Turing machine has access to some amount of random bits then is a modified EL Theorem such that with high probability, a simple program will produce a member of a set? This paper shows an EL Theorem for polynomial time constraints. A natural area of study would be over $\K^t$ with exponential $t$.  Is there one in which the universal Turing machine has space constraints?. In the recent literature, there are notions of time bound Kolmogorov complexity enhanced by random bits \cite{GoldbegKaLuOl22,Oliveira19}. Can they be used to create randomized, time bounded, EL Theorems? Given a new EL Theorem with constraints and/or random bits, are there accompanying derandomization theorems that can be proven?
\section{Conventions}
\label{sec:conv}
As noted in the introduction, $\K(x|y)$ is the conditional prefix free Kolmogorov complexity.  $\m(x)$ is the algorithmic probability. $\I(x;\ch)=\K(x)-\K(x|\ch)$ is the amount of information that the halting sequence $\ch\in\IS$ has about $x$. For some function $t:\N\rightarrow\N$, the $t$-time bounded Kolmogorov complexity is $\K^t(x) = \min\{\|p\|:U(p)=x\textrm{ in time }t(\|x\|)\}$. A probability is \textit{elementary}, if it has finite support and rational values. The deficiency of randomness of $x$ relative to a elementary probability measure $Q$ is $\d(x|Q)=-\log Q(x)-\K(x|Q)$. We recall for a set $D\subseteq\FS$, $\m(D)=\sum_{x\in D}\m(x)$. For the nonnegative real function $f$, we use $\lea f$, $\gea f$, and $\eqa f$ to denote $<f+O(1)$, $>f-O(1)$, and $=f\pm O(1)$. We also use $\lel f$ and $\gel f$ to denote $<f + O(\log (f+1))$ and $>f - O(\log (f+1))$, respectively. Derandomization of Section \ref{sec:channel} uses the following lemma, which is conservation of mutual information information with the halting sequence over deterministic processing.

\begin{lmm}[\cite{EpsteinDerandom22}]
\label{lmm:consH}
For partial computable $f:\FS\rightarrow\FS$, $\I(f(a);\ch)\lea\I(a;\ch)+\K(f)$.
\end{lmm}
\section{A New Proof to the EL Theorem}
\label{sec:el}

This section shows a new proof to the Sets Have Simple Members Theorem \cite{Levin16,Epstein19}. We also provide a proof that non-stochastic elements have high mutual information with the halting sequence, a well known result in the literature. This proof also does not rely on left-total machines, which the original proof did.

\begin{dff}[Stochasticisty]
A string $x$ is $(\alpha,\beta)$-stochastic if there exists an elementary probability measure $Q$ such that
$$\K(Q)\leq \alpha \textrm{ and }\d(x|Q)\leq \beta.$$
\end{dff}
\begin{thr}[Epstein,Levin]
\label{thr:el}
Let $P$ be a lower-semicomputable semimeasure and $c$ be a large constant. Every $(\alpha,\beta)$-stochastic set $D$ with $s=\ceil{-\log P(D)}$ contains an element $x$ with
$$\K(x)<  s + \alpha + 2\log \beta + \K(s) +2\log \K(s)+c.$$
\end{thr}
The theorem is directly implied by the following lemma.\newpage
\begin{lmm}
\label{lmm:el}
Let $P$ be a lower-semicomputable semimeasure and $c$ be a large constant. If a set $D$ is $(\alpha,\beta)$-stochastic relative to an integer $s=\ceil{-\log P(D)}$, then $D$ contains an element $x$ with
$$
\K(x) < s + \alpha + \log \beta + \K(\log \beta) + \K(s)+c.
$$
Note that if $y$ is $(\alpha,\beta)$-stochastic relative to $s$, then it is $(\alpha,\beta+\K(s))$-stochastic. Hence the lemma implies the theorem.
\end{lmm}
\begin{lmm}
\label{lmm}
Let $P$ be a discrete mesure and $Q$ be a measure on sets. There exists a set $S$ of size $\ceil{\beta/\gamma}$ such that
$$
Q(\{ D : P(D)\geq\gamma \textrm{ and $D$ is disjoint from }S\})\leq \exp (-\beta).
$$
\end{lmm}
\begin{prf}
We use the probabilistic method, and show that if we draw $\ceil{\beta/\gamma}$ elements according to the distribution $P$, then the obtained set $S$ satisfies the inequality with positive probability. The probability that a fixed set $D$ with $P(D)\geq\gamma$ is disjoint from $S$ is
$$\leq (1-\gamma)^{\beta/\gamma}\leq \exp(-\beta).$$
Hence the expected $Q$-measure of such a $D$ is at most $\exp(-\beta)$ and the required set $S$ exists.\qed
\end{prf}$ $\\

\noindent\textbf{Proof of Lemma \ref{lmm:el} for computable $P$}. Let $Q$ be an elementary probability measure with $\K(Q)\leq \alpha$ and $\d(D|Q,s)\leq \beta$. Without loss of generality, we assume that $\beta$ is large positive power of 2. Fix a search procedure that on input $Q$, $\beta$, and $\gamma=2^{-s}$ finds a set satisfying the conditions of Lemma \ref{lmm}.\\

For large $\beta$, the set $D$ must intersect the obtained set $S$. Indeed, consider the $Q$-test $g(X|Q,s)$ that is equal to $\exp(\beta)$ if $X$ is disjoint from $S$, and is zero otherwise. This is indeed a test, because the above lemma implies that its expected value for $X\sim Q$ is bounded by 1. Since the test is also computable, it is a lower bound to the optimal test $\t(X|Q,s)$, up to a constant factor. By stochasticity of the set $D$, $g(D|Q,s)<O(1)\t(D|Q,s)<O(2^{\beta})$, because $2^{\d(X|Q,s)}$ is an optimal $Q$ test relative to $s$. Thus for large enough $\beta$, $D$ intersects $Q$.\\

It remains to construct a description of each element in $S$ of the size given in the proposition. We construct a special decompressor that assigns short description to each element in $S$. On input of a string, the decompressor interprets the string as a concatenation of 4 parts:
\begin{enumerate}
\item A prefix-free description of $Q$ of size at most $\alpha$.
\item A prefix-free description of $\log \beta$ of size $\K(\log \beta)$.
\item A prefix-free description of $s$ of size $\K(s)$.
\item An integer of bitsize $\log (\beta/\gamma) = s + \log \beta$. 
\end{enumerate}
It interprets the last integer as the index of an element in the set $S$ of size $\ceil{\beta/\gamma}$ that is computed by the search procedure on input $Q$, $\beta$, and $\gamma$. The element is the output of the decompressor. The proposition is proven for computable $P$.\qed

\begin{rmk}
If $P$ is computable, a set $S$ satisfying the conditions of the lemma can be easily searched. But if $P$ is not computable, then the collection of sets $D$ with $P(D) \geq\gamma$ grows over time. Thus after constructing a good S, it can happen that a large $Q$-measure of sets $D$ appears that does not contain an element from $S$, and that new elements to $S$ need to be added. This type of interactive construction leads to an equivalent characterization of the problem in terms of a game which is shown in \cite{Shen12}. Below, another proof is presented. 
\end{rmk}
\noindent\textbf{Proof of Lemma \ref{lmm:el} for lower-semicomputable $P$}. We still assume that $\beta$ is a large power of 2. Let $\gamma=2^{-s}/2$. We can rewrite $P=\frac{\gamma}{\beta}(P_1+\dots+P_f+P_*)$, with $f\leq \beta/\gamma$, such that $P_1,\dots P_f$ are probability measures with finite support obtained by a lower semi-computable approximation of $P$, and $P_*$ is a lower-semicomputable semimeasure.\\

\noindent\textit{Construction of a lower-semicomputable test $g$ over sets.} We first construct tests $g_1,\dots g_f$ together with a list of strings $z_1,\dots,z_f$. Let $g_0(X)=1$. Assume we already constructed $z_1,\dots,z_{i-1}$ and $g_{i-1}$ for some $i=1,\dots,f$. Choose $z_i$ such that the test
$$
g_i(X) = 
\left\{
\begin{array}{ll}
g_{i-1}(X)& \textrm{if }g_{i-1}(X)\geq \exp(\beta)\\
\exp(P_i(X))g_{i-1}(X)& \textrm{if }g_{i-1}(X)<\exp(\beta)\textrm{ and $X$ is disjoint from }\{z_1,\dots,z_i\}\\
0& \textrm{otherwise.}\\
\end{array}
\right.
$$
satisfies $\E g_i(X) \leq \E g_{i-1}(X)$ where the expectations are taken for $X\sim Q$. Let $g(X)$ be equal to $\exp(\beta)$ if there exists an $i$ such that $g_i(X)\geq\exp\beta$, otherwise let $g(X)=0$. \textit{End of construction}\\

We first show that each required string $z_i$ in the construction exists. Suppose $z_1,\dots,z_{i-1}$ and $g_{i-1}$ have already been constructed. We show the existence of $z_i$ using the probabilistic method. If we draw $z_i$ according to $P_i$, then for each set $X$ for which the second condition of $g_i$ is satisfied, we have
$$\E_{z_i\sim P_i} g_i(X)\leq (1-P_i(X))g_{i-1}(X)\exp P_i(X) \leq g_{i-1}(X),$$
because of the inequality $1+r\leq \exp(r)$ for all reals $r$. If $X$ satisfies the first or third condition, then $\E g_i(X)\leq \E g_{i-1}(X)$ is trivially true. So
\begin{align*}
\E_{X\sim Q}\E_{z_i\sim P_i}g_i(X)&\leq \E_{X\sim Q}g_{i-1}(X),\\
\E_{z_i\sim P_i}\E_{X\sim Q}g_i(X)&\leq \E_{X\sim Q}g_{i-1}(X),
\end{align*}
and the required $z_i$ exists.

We have $G(x)\leq O(\t(X|Q,(\gamma,\beta)))$, where $\t$ is the optimal test because the construction implies $\E g\leq 1$ and is effective, thus $g$ is lower semicomputable. Every set $X$ with $P(X)\geq 2^{-s}=2\gamma$ satisfies $P_1(X)+\dots+P_f(X)\geq\frac{\beta}{\gamma}P(D)-1\geq 2\beta -1 \geq \beta$ by choice of $P_i$. Any such $X$ that is disjoint from the set $\{z_1,\dots,z_f\}$ satisfies
$$
g_f(X) = \exp(P_1(X))\exp(P_2(X))\dots \exp(P_f(X))\geq \exp(\beta).
$$
This implies $\d(X|Q,s)>\beta$ for large $\beta$, because up to $O(1)$ constants, we have
$$
1.44\beta\leq \log g(X)\leq \d(X|Q,(\beta,\gamma))\leq \d(X|Q,s) + 2\log \beta.
$$
By the assumption on $(\alpha,\beta)$-stochasticity of $D$, we have $\d(D|Q,s)\leq \beta$ and hence $D$ must contain some $z_j$. The theorem follows by constructing a description for each string $z_i$ of bitsize $s+\alpha +\log\beta +\K(\log \beta) + \K(s)$ in a similar way as above.\qed

\subsection{Non-Stochastic Objects}
The stochasticity of an object can be measured by 
$$\Ks(x) =\min\{\K(P)+O(\log\max\{ \d(x|P),1\}): P\textrm{ is an elementary probability measure}\}.$$
This term combines the complexity of the model $P$ with how well it fits $x$, i.e. the randomness deficiency $\d$. It is well known in the literature that non-stochastic objects have high mutual information with the halting sequence \cite{VereshchaginSh17}. In the following lemma, we reprove this fact, without using left-total machines, which was used in the original proof. 
\begin{lmm}
\label{lmm2}
$\Ks(x)\lel \I(x;\ch)$.
\end{lmm}
\begin{prf}
We dovetail all programs to the universal Turing machine $U$. For $p\in \mathrm{Domain}(U)$, $n(p)\in\N$ is the position in which the program $p\in\FS$ terminates. Let $\Omega^n = \sum_{p:n(p)< n}2^{-\|p\|}$ and $\Omega=\Omega^\infty$ be Chaitin's Omega. Let $\Omega^n_t$ be $\Omega^n$ restricted to the first $t$ digits. Let $x^*\in\BT^{\K(x)}$, with $U(x^*)=x$ with minimum $n(x^*)$. Let $k(p)=\max\{\ell:\Omega^{n(p)}_{\ell}=\Omega_{\ell}\}$ and $k=k(x^*)$. We define the elementary probability measure $Q(x) = \max\{2^{-\|p\|+k} : k(p)=k, U(p)=x\}$, $Q(\emptyset)=1-Q(\FS\setminus\{\emptyset\})$.
\begin{align*}
\d(x|Q) &= -\log Q(x) - \K(x|Q)\lea (\K(x) - k)- \K(x|\Omega_k)\\
         &\lea (\K(x|\Omega_k) + \K(\Omega_k) - k) - \K(x|\Omega_k)\lea (k + \K(k)) - k\\
          &\lea \K(k).\\ \\
\K(x|\ch) &\lea\ \K(x|Q) + \K(Q|\ch)\lea  \K(x|Q) + \K(\Omega_k|\ch)\\
&\lea -\log Q(x) + \K(k)\lea(\K(x) - k) + \K(k)\\
k & \lel \K(x) - \K(x|\ch)\\ \\
\Ks(x) &\lea \K(Q) + O(\log\max\{ \d(x|P),1\})\lea k + O(\K(k))\lel\I(x;\ch).
\end{align*}\qed
\end{prf}$ $\\

The following corollary comes from Theorem \ref{thr:el} and Lemma \ref{lmm2}. 
\begin{cor}[Epstein,Levin]
\label{cor:el}
For finite $D\subset\FS$, $\min_{x\in D}\K(x) \lel -\log \m(D) + \I(D;\ch)$.
\end{cor}
\section{Classical Channels}
\label{sec:channel}
There are deep connections between classical information theory and algorithmic information theory, with many theorems of the former appearing in an algorithmic form in the latter. In this section we revisit this connection. In particular we prove properties about the compression size of shared codebooks.  A standard setup in information theory is two parties Alice and Bob who want to communicate over a noisy channel and share a codebook over a noiseless channel. However one might ask is how many bits did it take to communicate the codebook? By using derandomization, the tradeoff between codebook complexity and communication capacity can be proven.\newpage

\begin{dff}[Discrete Memoryless Channel]
 The input and output alphabets $\mathcal{X}$ and $\mathcal{Y}$ are finite. The channel $(\mathcal{X},p(y|x),\mathcal{Y})$ is represented by a conditional probability distribution $p(y|x)$. To send multiple symbols, we have $p(y^n|x^n)=\prod_{i=1}^np(y_i|x_i)$. The capacity of channel with respect to a distribution $Q$ over $\mathcal{X}$ is
 $$ C_Q = I(X:Y)\textrm{ where random variables $(X,Y)$ are distributed according to }Q(x)p(y|x).$$
 The term $I$ is the mutual information between random variables.
\end{dff}

\begin{dff}[Codebook]
A $(M,n)$ codebook for channel $(\mathcal{X},p(y|x),\mathcal{Y})$ contains the following:
\begin{enumerate}
\item An encoder $\mathrm{Enc}_n:\{1,\dots,M\}\rightarrow\mathcal{X}^n$.
\item A decoder $\mathrm{Dec}_n:\mathcal{Y}^n\rightarrow \{1,\dots,M\}$.
\end{enumerate}
The \textit{rate} of the codebook is $R=\frac{\log M}{n}$. The conditional probability of error is $\lambda_i=\sum_{y_n} p(y^n|x^n=\mathrm{Enc}(i))[\mathrm{Dec}(y^n)\neq i]$, where $[\cdot]$ is the indicator function. The average error rate of the codebook with respect to a fixed channel $p$ is $P^{(n)}_e = \frac{1}{M}\sum_{i=1}^M\lambda_i$. It is the probability that, given the uniform distribution over $\{1,\dots,M\}$ for the sending symbols, the receiver decodes a symbol different from the encoded one.
\end{dff}

This section shows the following high level description of a communication scheme is possible: there is a sender Alice and a receiver  Bob that communicate through a noisy memoryless discrete channel and Alice can send a codebook to Bob once on a side noiseless channel. Bob has oracle acess to the channel function $p(y|x)$ but Alice does not. Given a computable distribution $Q$ over the input alphabet, and assuming the channel is non-exotic, Alice can hypothetically send ${\sim}\K(Q)$ bits plus some encoded parameters describing a codebook to Bob on the side channel. Then Alice and Bob can communicate with any rate $R$ less than the capacity $C_Q$ over the noisy channel. This setup is formalized with Theorem \ref{thr:channel}. To prove this theorem, some results are needed from classical information theory.
\subsection{Jointly Typical Sequences}
We need the following definition and theorem, which can be found in \cite{CoverJo91}, in the proof of Theorem \ref{thr:channel}. $H(X)$ is the entropy of random variable $X$, and $I(X:Y)$ is the mutual information between random variables $X$ and $Y$.
\begin{dff}
The set $A_\epsilon^{(n)}$ of jointly typical sequences $\{(x^n,y^n)\}$ with respect to the distribution $p(x,y)$ is the set of $n$-sequences with empirical entropies $\epsilon$-close to the true entropies. $\mathcal{X}$ and $\mathcal{Y}$ are the finite discrete alphabet of random variables $X$ and $Y$. Let $p(x^n,y^n)=\prod_{i=1}^np(x_i,y_i)$.
\begin{align*}
A_\epsilon^{(n)} =& \{ (x^n,y^n)\in \mathcal{X}^n\times \mathcal{Y}^n:\\
&\left|-\frac{1}{n}\log p(x^n)-H(X)\right|<\epsilon,\\
&\left|-\frac{1}{n}\log p(y^n)-H(Y)\right|<\epsilon,\\
&\left.\left|-\frac{1}{n}\log p(x^n,y^n)-H(X,Y)\right|<\epsilon\right\}.\\
\end{align*}
\end{dff}
The following theorem details properties about the set $A^{(n)}_\epsilon$. A proof for it can be found in \cite{CoverJo91}.
\begin{thr}[Joint AEP]
\label{thr:jaep}
Let $(X^n,Y^n)$ be sequences of length $n$ drawn i.i.d. according to $p(x^n,y^n)=\prod_{i=1}^np(x_i,y_i)$. Then
\begin{enumerate}
\item $\Pr\left((X^n,Y^n)\in A^{(n)}_\epsilon\right)\rightarrow 1 - o(1)$.
\item If $(\tilde{X}^n,\tilde{Y}^n)\sim p(x^n)p(y^n) (\textrm{$\tilde{X}^n$ and $\tilde{Y}^n$ are independent with the same marginals as $p(x^n,y^n)$})$, then
$\Pr\left((\tilde{X}^n,\tilde{Y}^n)\in A^{(n)}_\epsilon\right)
\leq 2^{-nI(X:Y)-3\epsilon}$.

\end{enumerate}
\end{thr}
\subsection{Naive Sender Paradigm}
\begin{thr}
\label{thr:channel}
For channel $\mathfrak{C}=(\mathcal{X},p(y|x),\mathcal{Y})$ and every computable distribution $Q$ over $\mathcal{X}$, for every rate $R<C_Q$, there is a $(2^{nR},n)$ codebook $(\mathrm{Enc}_n,\mathrm{Dec}_n)$ with rate $R$ and average error rate $o(1)$ such that there is a program $p$ with $\|p\|\lel\K(n,R,Q)+\I((n,R,Q,\mathfrak{C});\ch)$ and
\begin{align*}
U(p,x)&=\mathrm{Enc}_n(x),\\
U(p,\mathfrak{C},x)&=\mathrm{Dec}_n(x).
\end{align*}
\end{thr}

\begin{prf}

We start by generating a $(2^{nR},n)$ code randomly according to distribution $Q$. We generate $2^{nR}$ codewords $x\in\mathcal{X}$ independently according to the distribution
$$
Q(x^n) = \prod_{i=1}^np(x_i).
$$
The codewords can be represented as rows of a matrix
$$
\mathcal{C} = 
\begin{bmatrix}
x_1(1) & x_2(1)&\dots&x_n(1)\\
\vdots&\vdots&\ddots&\vdots\\
x_1(2^{nR}) & x_2(2^{nR})&\dots& x_n(2^{nR})\\
\end{bmatrix}
$$
Each entry is generated i.i.d according to $Q(x)$, with
$$
\Pr(\mathcal{C}) = \prod_{w=1}^{2^{nR}}\prod_{i=1}^n p(x_i(w)).
$$
Consider the following algorithm for encoding and decoding a message.
\begin{enumerate}
\item A random code $\mathcal{C}$ is generated according to $Q(x)$.
\item The code $\mathcal{C}$ is sent to both the sender and the receiver. \textit{Only the receiver is assumed to know the channel transition matrix $p(y|x)$ for the channel}. This differs from the standard literature, which assumes knowledge of $p$ by the sender.
\item A message $W$ is chosen according to the uniform distribution. 
$$
\Pr(W=w)=2^{-nR},\hspace*{1.5cm} w=1,2,\dots,2^{nR}.
$$ 
\item The $w$th codeword $X^n(w)$ corresponding to the $w$th row of $\mathcal{C}$ is sent over the channel.
\item The receiver receives a sequence $Y^n$ according to the distribution
$$
P(y^n|x^n(w))=\prod_{i=1}^np(y_i|x_i(w)).
$$
\item The receiver decares that the index $\hat{W}$ was sent if the following conditions are satisfied:
\begin{itemize}
\item $(X^n(\hat{W}),Y^n)$ is jointly typical, i.e. $(X^n(\hat{W}),Y^n)\in A^{(n)}_\epsilon$.
\item There is no other index $W'\neq \hat{W}$ such that $(X^n(W'),Y^n)\in A^{(n)}_\epsilon$.
\end{itemize}
If no such $\hat{W}$ exists or if there are more than one, an error is declared, and the decoder outputs 0.
\item There is a decoding error if $\hat{W}\neq W$. Let $\mathcal{E}$ be this event.
\end{enumerate}
We now analyze the probability of the error with respect to the random codebook $\mathcal{C}$.
\begin{align}
\nonumber
\Pr(\mathcal{E}) &= \sum_\mathcal{C}\Pr(\mathcal{C})P_e^{(n)}(\mathcal{C})\\
\nonumber
&=\sum_{\mathcal{C}}\Pr(\mathcal{C})\frac{1}{2^{nR}}\sum_{w=1}^{2^{nR}}\lambda_w(\mathcal{C})\\
\nonumber
&=\frac{1}{2^{nR}}\sum_{w=1}^{2^{nR}}\sum_\mathcal{C}\Pr(\mathcal{C})\lambda_w(\mathcal{C})\\
\label{eq:er1}
&=\sum_\mathcal{C}\Pr(\mathcal{C})\lambda_1(\mathcal{C})\\
\nonumber
&=\Pr(\mathcal{E}|W=1),
\end{align}
where Equation \ref{eq:er1} is due to symmetry of the code construction. We define 
$$
E_i = \{(X^n(i),Y^n)\in A^{(n)}_\epsilon\},\hspace*{1cm} i\in\{1,2,\dots,2^{nR}\}.
$$
So $E_i$ is the event that the $i$th code and $Y^n$ are jointly typical, noting that $Y^n$ is the result of sending the first codeword $X^n(1)$ over the channel. So
\begin{align*}
\Pr(\mathcal{E}|W=1) &= P(E_1^c\cup E_2\cup E_3\dots E_{2^{nR}}|W=1)\leq P(E^c_1|W=1)+\sum_{i=2}^{2^{nR}}P(E_i|W=1).
\end{align*}
Due to the code generation procedure, $X^n(1)$ and $X^n(i)$ are independent for $i\neq 1$, and therefore, so are $Y^n$ and $X^n(i)$. Due to Theorem \ref{thr:jaep} (2), the probability that $X^n(i)$ and $Y^n$ are jointly typical is $\leq 2^{-n(I(X;Y)-3\epsilon)}$, where random variables $X$ and $Y$ are distributed acording to $Q(x)p(y|x)$. So by Theorem \ref{thr:jaep} (1), for sufficiently large $n$,
\begin{align*}
\Pr(\mathcal{E}) &=\Pr(\mathcal{E}|W=1) \leq P(E^c_1|W=1)+\sum_{i=2}^{2^{nR}}P(E_i|W=1)\\
&\leq\epsilon+\sum_{i=2}^{2^{nR}}2^{-n(I(X:Y)-3\epsilon)}\\
&=\epsilon +\left(2^{nR-1}\right)2^{-n(I(X:Y)-3\epsilon)}\\
&\leq \epsilon+ 2^{3n\epsilon}2^{-n(I(X:Y)-R)}\\
&\leq 2\epsilon,
\end{align*}
under the condition $R<I(X:Y)-3\epsilon=C_Q-3\epsilon$. Hence if $R<C_Q$ we can choose an $\epsilon$ and $n$ so the average probability of error, averaged over codebooks is less than $2\epsilon$. We now remove the average over codebooks. Since the average error rate $P_e(\mathcal{C})$ is small, there exists at least one codebook $\mathcal{C}^*$ with a small average probability of error, with
\begin{align*}
\Pr(\mathcal{E}|\mathcal{C}*) &= \frac{1}{2^{nR}}\sum_{i=1}^{2^{Rn}}\lambda_i(\mathcal{C}^*)\leq 2\epsilon.
\end{align*}

\noindent\textbf{Connection with Algorithmic Information Theory.} We now derive the statements of the theorem. Define $P$ to be the probability over codebooks used in earlier in this proof that uses the distribution $Q$ to generate the codewords. Thus $\K(P)\lea \K(Q,n,R)$. Let $D$ be the set of encoded codebooks that achieve an error rate less than or equal to $2\epsilon$. By the arguments above, $P(D)\geq 0.5$. This set $D$ is computable from $Q$, $n$, $R$, and $\mathfrak{C}$, with $\K(D|(Q,n,R,\mathfrak{C}))=O(1)$. Thus by Corollary \ref{cor:el} and Lemma \ref{lmm:consH}, there is a codebook $\mathcal{C}^*\in D$ that has an error rate $\leq 2\epsilon$, with
\begin{align}
\nonumber
\K(\mathcal{C}^*)&\lel\K(P) -\log P(D)+\I(D;\ch)\\
\label{eq:CStar}
&\lel\K(Q,n,R) + \I((Q,r,n,\mathfrak{C});\ch).
\end{align}
Thus the sender can use solely $\mathcal{C}^*$ to send messages to the receiver. The receiver needs to determine if sequences are jointly typical, and thus uses $(\mathcal{C}^*,Q,\mathfrak{C})$ to decode the messages. Note that with careful analysis of the proof of Lemma \ref{lmm:el} for computable probabilities, one can construct a short program for $\mathcal{C}^*$ (with size less than that of Equation \ref{eq:CStar}) that can also compute $Q$. Thus, we can construct a program $p$ with the properties described in the theorem statement.
\qed

\end{prf}
\section{Resource Bounded EL Theorem}

In this section we derive the resource bounded EL theorem. We also derive an interesting corollary to Theorem 4.1 in \cite{AntunesFo09} which states to invert a hash function $f^{-1}(x)$, one can find a secret key $\pi$ of size approximiately equal to $x$ that will efficiently decompress to a  pre-image of $x$ with respect to $f$. The results in this section are not unconditional, they require the existence of the pseudorandom generator, introduced in \cite{NisanWi94}.
\begin{asm}
\label{asm}
\textbf{Crypto} is the assumption that there exists a language in $\mathbf{DTIME}(2^{O(n)})$ that does not have size $2^{o(n)}$ circuits with $\Sigma_2^p$ gates. This asssumption is need in the proof of Theorem \ref{thr:samp} in \cite{AntunesFo09} to assume the existence of a pseudorandom generator $g:\BT^{k\log n}\rightarrow\BT^n$, computable in time polynomial in $n$.
\end{asm}
\begin{dff}
$\mathbf{FP}'=\{f:f\in\mathbf{FP}\textrm{ and }\|x\|=\|f(x)\|\}$.
\end{dff}
\begin{dff}
For $A\in\mathbf{FP}'$ we say that $A$ samples $D\subset\BT^n$ with probability $\gamma$, if $|\BT^n\cap A^{-1}(D)|/2^n>\gamma$.
\end{dff}

\begin{thr}[\cite{AntunesFo09}]
\label{thr:samp}
Assume \textbf{Crypto}. Let $F\in\mathbf{FP}'$. Let $m,n\in\N$ where $\BT^n\supseteq f(\BT^m)$. Let $T_y = \{w\in\BT^m:F(w)=y\}$ and $V_k=\{y:\|y\|=n\textrm{ and }
|T_y|\geq 2^k\}$. There exists a function 
$$G:\Sigma^{m-k+O(\log m)}\rightarrow \Sigma^m$$ computable in polynomial time  such that for all $y\in V_k$, $\mathrm{range}(G)\cap T_y\neq\emptyset$.
\end{thr}
\begin{rmk}
 In the previous theorem, the running time of $G$ is a polynomial function of the running time of $F$. This was noted in \cite{LuOlZi22}.
In addition, in subsequent theorems and corollaries of this section, the polynomial time function $p$ in the resource bounded complexity $\K^p$ is a polynomial function of the running times of the algorithms of the theorem/corollary statements. Furthermore, due to \cite{AntunesFo09}, $G$ can be encoded in $O(1)$ bits.
 \end{rmk}
The following corollary implies that to invert $x$ with a hash function $f$, one can find a secret key $\pi$ of size approximately equal to $x$ that efficiently expands to an element in $f^{-1}(x)$.
\begin{cor}
Assume \textbf{Crypto}. Let $f\in\mathbf{FP}'$, where $f(\BT^n)\subseteq\BT^{n-k}$. Then for some polynomial $p$ where for $\BT^n\supseteq D= f^{-1}(x)$,
$$\min_{y\in D}\mathbf{K}^p(y) = n - \log |D|+ O(\log n).$$
\end{cor}
\begin{prf}
Follows directly from Theorem \ref{thr:samp}.
\qed
\end{prf}
\begin{cor}[Resource EL]
\label{cor:resel}
Assume \textbf{Crypto}. Let $L\in\mathbf{P}$, $A\in\mathbf{FP}'$, and assume $A$ samples $L_n$ with probability $\delta_n$. Then for some polynomial $p$,
$$\min_{x\in L_n}\mathbf{K}^p(x) < -\log\delta_n+ O(\log n).$$
\end{cor}
\begin{prf}
Let $F\in\mathbf{FP}'$ where $F(\BT^n)\subseteq \BT^n$ and for $x\in\BT^n$, $F(x)=1^n$ if $A(x)\in L_n$ and $F(x)=0^n$ otherwise. Let $k\in \N$ be maximal such that $\delta_n\geq 2^{k-n}$. Let $\ell=n-k+O(c\log n)$.  By Theorem \ref{thr:samp}, there exists a function $G:\BT^{\ell}\rightarrow\BT^n$ running in polynomial time such that there exists $x\in \ell$, with $G(x)=1^n$. This is because $1^n\in T_k$, using the definition in Theorem \ref{thr:samp}, because $A$ produces a member of $L_n$ with probability at least $\delta_n$ and all of $L_n$ is mapped to $1^n$. We define a program $P$ that uses $G$ to map $x$ to a string $y$, then use $A$ to map $y$ to a string $z\in L_n$. This program $P$ is of size $\ell$ and runs in polynomial time.
\qed
\end{prf}$ $\\

A verifier $V:\FS\times\FS\rightarrow\BT$ is a function computable in polynomial time with respect to the first argument. For a given $x$, $\mathrm{Proofs}(x) = \{ y : V(x,y)=1\}$.
\begin{cor}
\label{cor:verifier}
Assume \textbf{Crypto}. Let $\{x_n\}$ be uniformly computable in polynomial time. For a verifier $V(x,y)$, let $A\in\mathbf{FP}'$ sample $\mathrm{Proofs}(x_n)$ with probability $\gamma_n$. Thus there is a polynomial $p$ and $y\in \mathrm{Proofs}(x_n)$ with
$$\mathbf{K}^p(y)< -\log \gamma_n+O(\log n).$$
\end{cor}
\section{Resource Bounded Derandomization}
In this section, we use Corollary \ref{cor:verifier} to produce three examples of resource bounded derandomization. The resource free versions of these theorems can be found in \cite{EpsteinExmDerand22}.
\begin{lmm}[Lovasz Local Lemma]
\label{lmm:lll}
Let $E_1,\dots, E_n$ be a collection of events such that $\forall i:\Pr[E_i]\leq p$. Suppose further that each event is dependent on at most $d$ other events, and that $ep(d + 1) \leq 1$. Then, $\Pr\left[\bigcap_i\overline{E}_i\right]>\left(1-\frac{1}{d+1}\right)^n$.
\end{lmm}

\begin{prp}[Mutual Independence Principle]
\label{prp:mip}
Suppose that $Z_1,\dots Z_m$ is an underlying sequence of independent events and suppose that each event $A_i$ is completely determined by some subset $S_i\subset\{Z_1,\dots,Z_m\}$. If $S_i\cap S_j=\emptyset$ for $j=j_1,\dots,j_k$ then $A_i$ is mutually independent of $\{A_{j_1},\dots,A_{j_k}\}$.
\end{prp}$ $\newpage

\subsection{{\sc Vertex-Disjoint-Cycles}}

\begin{thr}
Assume \textbf{Crypto.} Let $\{G_n\}$ be a uniformly computable in polynomial time sequence of $k$-regular graphs, with $k\geq 5$. There is a polynomial $p$ where for each $G_n$, there is a partition $x$ of $\lfloor\frac{k}{3\ln k}\rfloor$ components each containing a cycle with
$$
\mathbf{K}^p(x) < 2n/k^2+O(\log n).
$$
\end{thr}

\begin{prf}
We partition the vertices of $G$ into $c=\floor{k/3\ln k}$ components by assigning each vertex to a component chosen independently and uniformly at random. With positive probability, we show that every component contains a cycle. It is sufficient to prove that every vertex has an edge leading to another vertex in the same component. This implies that starting at any vertex there exists a path of arbitrary length that does not leave the component of the vertex, so a sufficiently long path must include a cycle. A bad event $A_v=\{\textrm{vertex $v$ has no neighbor in the same component}\}$. Thus
\begin{align*}
\Pr[A_v] &=\prod_{(u,v)\in E}\Pr[\textrm{$u$ and $v$ are in different components}]\\
&=\left(1-\frac{1}{c}\right)^k<e^{-k/c}\leq e^{-3\ln k}=k^{-3}.
\end{align*}
$A_v$ is determined by the component choices of itself and of its out neighbors $N^{\mathrm{out}}(v)$ and these choices are independent. Thus by the Mutual Independence Principle, (Proposition \ref{prp:mip}) the dependency set of $A_v$ consist of those $u$ that share a neighor with $v$, i.e., those $u$ for which $(\{v\}\cup N(v))\cap(\{u\}\cup N(u))\neq 0$. Thus the size of this dependency is at most $d=(k+1)^2$.

Take $d=(k+1)^2$ and $p=k^{-3}$, so $ep(d+1)=e(1+(k+1)^2)/k^3\leq 1$, holds for  $k\geq 5$. Thus, noting that $k\geq 5$, by Lovasz Local Lemma, (Lemma \ref{lmm:lll}),
\begin{align}
\label{eq:hyper3}
\Pr\left[\bigcap_{v\in G}\overline{A}_v\right]&>\left(1-\frac{1}{d+1}\right)^n=\left(1-\frac{1}{(k+1)^2+1}\right)^n>\left(1-\frac{1}{k^2}\right)^n.
\end{align}
Graphs $G_n$ of size $n$ are encoded in strings of size $kn\ceil{\log n}$ and partitions are the proofs, encoded in strings of size $n\ceil{\log k}$. The verify $V$ returns 1 if each partition contains a cycle. The verifier runs in time $O(n\log n)$. We define a sampling function $A\in\mathbf{FP}'$ over the partition/proofs that is the same as the probability used in the Lovasz Local Lemma, i.e. the uniform distribution. Thus $A(x)=x$. $A$ samples Proofs($G_n$) with probability $\gamma_n$, where by Equation \ref{eq:hyper3},
$$-\log \gamma_n< -n\log(1-1/k^2) < 2n/k^2.$$
Thus by Corolloray \ref{cor:verifier}, there is a polynomial $p$, where for each graph $G_n\in Q$ of $n$ vertices, there is a partition $x\in\mathrm{Proofs}(G_n)$ with
$$
\K^p(x) < 2n/k^2+O(\log n).
$$

\qed
\end{prf}$ $\newpage

\subsection{{\sc Balancing-Vectors}}

\begin{cor} Assume \textbf{Crypto}. For vector $v$, $\|v\|_\infty =\max_i |v_i|$. A binary matrix $M$ has entries of 0s or 1s. Let $\{M_n\}$ be a uniformly polynomial time computable sequence of $n\times n$ binary matrices. There is a polynomial $p$ where for each $M_n$ there is a vector $b\in \{-1,1\}^n$ such that $\|M_nb\|_\infty\leq 4\sqrt{n\ln n}$ and
$$
\K^p(b) = O(\log n).
$$
\end{cor} 

\begin{prf}
Let $v=(v_1,\dots,v_n)$ be a row of $M$. Choose a random $b=(b_1,\dots,b_n)\in\{-1,+1\}^n$. Let $i_1,\dots,i_m$ be the indices such that $v_{i_j}=1$. Thus
$$
Y=\langle v,b\rangle=\sum_{i=1}^nv_ib_i=\sum_{j=1}^mv_{i_j}b_{i_j}=\sum_{j=1}^mb_{i_j}.
$$
$$
\E[Y]=\E[\langle v,b\rangle]=\E\left[\sum_iv_ib_i\right]=\sum_i\E[v_ib_i]=\sum v_i\E[b_i]=0.
$$
By the Chernoff inequality and the symmetry $Y$, for $\tau=4\sqrt{n\ln n}$,
$$\Pr[|Y|\geq\tau]=2\Pr[v\cdot b\geq \tau]=2\Pr\left[\sum_{j=1}^mb_{i_j}\geq \tau\right]\leq 2\exp\left(-\frac{\tau^2}{2m}\right)=2\exp\left(-8\frac{n\ln n}{m}\right)\leq 2n^{-8}.
$$
Thus, the probability that any entry in $Mb$ exceeds $4\sqrt{n\ln n}$ is smaller than $2n^{-8}$. Thus, with probability $1-2n^{-7}$, all the entries of $Mb$ have value smaller than $4\sqrt{n\ln n}$.

Let $A(x)=x$ be the uniform sampling function. The verifier $V$ takes in a matrix $M$ and a vector $b$ and returns 1 iff $\|Mb\|_\infty\leq 4\sqrt{n\ln n}$. Let $D\subset\BT^n$ consist of all strings that encode vectors $b_x\in\{-1,+1\}^n$ in the natural way such that $\|Mb_x\|_{\infty}\leq 4\sqrt{n\ln n}$. By the above reasoning, $A$ samples $D$ with probability $\geq 1- 2n^{-7} >0.5$. So by Corollary \ref{cor:verifier}, there is a polynomial $p$, where for each $n\times n$ matrix $M_n$ there is a binary vector $b\in\{-1,1\}^n$ with $\|Mb\|_\infty\leq 4\sqrt{n\ln n}$ and
$$
\K^p(b)=O(\log n).
$$
\qed
\end{prf}

\subsection{{\sc k-Sat}}

\begin{cor}
Assume \textbf{Crypto}. Let $\Phi_n$ be a $k(n)$-SAT formula, using $n$ variables, $m(n)$ clauses, uniformly polynomial time computable in $n$. Furthermore, each variable occurs in at most $2^{k(n)}/k(n)e-1$ clauses. There is a polynomial $p$ and a satisfying assignment $x$ of $\Phi_n$ where $$\mathbf{K}^p(x) < 2m(n)e2^{-k(n)} + O(\log n).$$
\end{cor}

\begin{prf}
 The sample space is the set of all $2^n$ assigments. We choose a random assignment, where each variable is independently equally likely to have a true or false assignment. For each clause $C_J$, $E_j$ is the bad event ``$C_j$ is not satisfied''. Let $p=2^{-k(n)}$ and $d=(2^{k(n)}/e)-1$. Thus $\forall j$, $\Pr[E_j]\leq p$ as each clause has size $k(n)$ and each $E_j$ is dependent on at most $d$ other events since each variable appears in at most $2^{k(n)}/k(n)e-1$ other clauses, and each clause has $k(n)$ variables. Thus since $ep(d+1)\leq 1$, by the Lovasz Local Lemma \ref{lmm:lll}, we have that, 
\begin{align}
\label{eq:ksat}
\Pr\left[\bigcap_j\overline{E_j}\right]>\left(1-\frac{1}{d+1}\right)^{m(n)}=\left(1-\frac{e}{2^{k(n)}}\right)^{m(n)}.
\end{align}
Let $D_n\subset\BT^n$ be the set of all assignments that satisfy $\phi_n$. We use a uniform sampler, with $A(x)=x$. By the above reasoning, $A$ samples $D_n$ with probability $\gamma_n>\left(1-\frac{e}{2^{k(n)}}\right)^{m(n)}$. Thus 
$$-\log \gamma_n < -m(n)\log \left(1-e/2^{k(n)}\right) < 2 em(n)2^{-k(n)}.$$
By Corollary \ref{cor:verifier}, there is a polynomial $p$, where for all $n$, there is a satisfying assignment $x\in D_n$ of $\Phi(n)$ with 
$$
\K^{p}(x) < 2m(n)e2^{-k(n)}+O(\log n).
$$
\qed
\end{prf}


\end{document}